\documentclass{appolb}
\usepackage{graphicx}
\usepackage{amssymb}

\begin{document}

\title{Charm and bottom quark production in photon-nucleon \\
and photon-photon collisions }

   \author{A. Szczurek$^{1},^{2}$  \\
   {\it $^{1}$ Institute of Nuclear Physics, PL-31-342 Cracow, Poland  } \\
   {\it $^{2}$ University of Rzesz\'ow, PL-35-959 Rzesz\'ow, Poland } \\ }
\maketitle

\begin{abstract}
I discuss mechanisms of heavy quark production
in (real) photon-nucleon and (real) photon - (real) photon
collisions. In particular, I focuse on application of
the Saturation Model.
In addition to the main dipole-nucleon or dipole-dipole contribution
included
in recent analyses, I propose how to calculate within
the same formalism the hadronic single-resolved contribution
to heavy quark production.
At high photon-photon energies this yields a sizeable correction of about
30-40 \% for inclusive charm production and 15-20 \%
for bottom production.
%
\end{abstract}

\section{Introduction}

The total cross section for virtual photon - proton
scattering in the region of small $x$ and intermediate $Q^2$
can be well described by the Saturation Model (SM)
\cite{GW98}.
The very good agreement with experimental data can be extended
even to the region of rather small $Q^2$ by adjusting
an effective quark mass.
At present there is no deep understanding of the fit value
of the parameter as we do not understand in detail the confinement
and the underlying nonperturbative effects related to large size
QCD contributions.

In this presentation I shall
limit to the production of heavy quarks which is
simpler and more transparent for real photons.
Here one can partially avoid the problem of the poor understanding
of the effective light quark mass, i.e. the domain of the large
(transverse) size of the hadronic system emerging from the photon.

It was shown recently that the simple SM description
can be succesfully extended also to the photon-photon
scattering \cite{TKM01}.
The heavy quark production in photon-photon collisions is interesting
in the context
of a deficit of standard QCD predictions relative
to the experimental data as observed recently for $b$ quark
production.

\section{Heavy quark production in photon-nucleon scattering}

In the picture of dipole scattering the cross section for
heavy quark-antiquark ($Q \bar Q$) photoproduction on
the nucleon can be written as
\begin{equation}
\sigma_{\gamma N \rightarrow Q \bar Q}(W) =
\int d^2 \rho d z \; |\Phi_T^{Q \bar Q}(\vec{\rho},z)|^2
\sigma_{d N}(\rho,z,W) \; ,
\label{SM_gamma_nuc}
\end{equation}
where $\Phi_T$ is (transverse) quark-antiquark photon wave
function (see for instance \cite{NZ91}) and
$\sigma_{dN}$ is the dipole-nucleon total cross section.
Inspired by its phenomenological success
\cite{GW98} we shall use the SM parametrization
for $\sigma_{dN}$.
Because for real photoproduction the Bjorken-$x$ is not defined
we are forced to replace $x$ by $x_g$
\cite{szczurek}.

In Fig.1a we show predictions of SM for charm photoproduction.
The dotted line represents calculations based on Eq.(\ref{SM_gamma_nuc}).
The result of this calculation exceeds considerably
the fixed target experimental data.
One should remember, however, that the simple formula (\ref{SM_gamma_nuc})
applies at high energies only. At lower energies one should
include effects due to kinematical threshold.
In the momentum representation this can be done by requiring:
$M_{Q \bar Q} < W$, where $M_{Q \bar Q}$ is the invariant mass of the
final $Q \bar Q$ system.
This upper limit
still exceeds the low energy experimental data.
There are phase space limitations in the region
$x_g \rightarrow$ 1 which have been neglected so far.
Those can be estimated using naive counting rules.
Such a procedure leads to a reasonable agreement with the
fixed target experimental data.

The deviation of the solid line from the dotted line gives an idea
of the range of the safe applicability of SM for
the production of the charm quarks/antiquarks.
The cross section for $W >$ 20 GeV is practically independent
of the approximate treatment of the threshold effects.
SM seems to slightly underestimate the H1 collaboration data \cite{H1_ccbar}.
For comparison in Fig.1 we show the result of
similar calculations in the collinear approach (thick dash-dotted line)
with details described in \cite{szczurek}.

The calculation above is not complete.
For real photons a vector dominance contribution due to
photon fluctuation into vector mesons should
be included on the top of the dipole contribution.
In the present calculation we include only the dominant
gluon-gluon fusion component. Then
\begin{equation}
\sigma_{\gamma N \rightarrow Q \bar Q}^{VDM}(W) =
\sum_V \frac{4 \pi}{f_V^2} \; \int dx_V dx_N \;
g_V(x_V,\mu^2_F) \; g_N(x_N,\mu^2_F) \; \sigma_{gg \rightarrow Q \bar
  Q}(\hat W) \; .
\label{gp_VDM}
\end{equation}
Here the $f_V$ constants describe the transition of the photon
into vector mesons ($\rho$, $\omega$, $\phi$).
The gluon distributions in vector mesons are taken as that
for the pion \cite{GRV_pion}.

The dash-dotted line in Fig.1a shows the VDM contribution
calculated in the leading order (LO) approximation for
$\sigma_{gg \rightarrow Q \bar Q}$.
The so-calculated VDM contribution cannot be neglected at high
energies.

The situation for bottom photoproduction seems similar.
In Fig.1b we compare the SM predictions
with the data from the H1 collaboration \cite{H1_bbbar}.
Here the threshold effects may survive up to very high energy
$W \sim$ 50 GeV.
Again the predictions of SM are slightly below
the H1 experimental data point. The relative magnitude
of the VDM component is similar as for the charm production.

\vspace{-0.1cm}
\begin{figure}
  \begin{center}
\hspace{-1.4cm}
    \includegraphics[width=5.9cm]{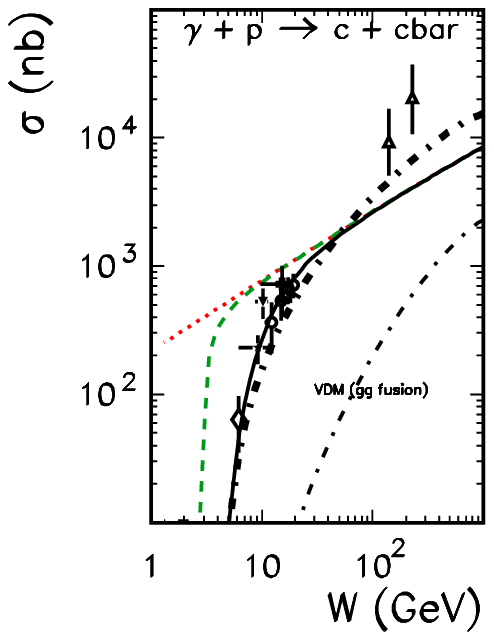}
\hspace{0.0cm}
    \includegraphics[width=5.9cm]{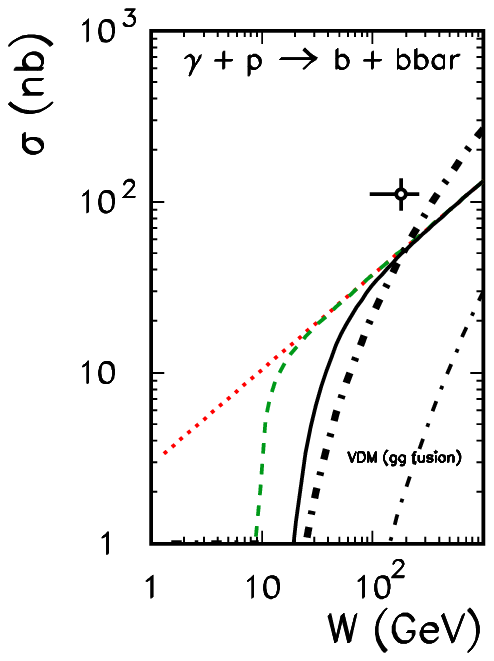}
  \end{center}
\vspace{-1.3cm}
\caption{\small
The cross section for $\gamma + p \rightarrow Q \bar Q X$.
The dotted line: standard SM,
the dashed line: includes kinematical threshold,
the solid line includes in addition a suppression by
$(1-x_c)^7$, the thick dash-dotted line:
collinear approximation and the thin dash-dotted line:
the LO VDM contribution.
}
\label{figure1}
\end{figure}

\section{Heavy quark production in photon-photon scattering}

In the dipole-dipole approach 
%
\begin{eqnarray}
\sigma_{\gamma \gamma \rightarrow Q \bar Q}^{dd}(W) &=&
\sum_{f_2 \ne Q} \int
                 |\Phi^{Q \bar Q}(\rho_1, z_1)|^2
                 |\Phi^{f_2 \bar f_2}(\rho_2, z_2)|^2
         \sigma_{dd}(\rho_1,\rho_2,x_{Qf}) \;
d^2 \rho_1 d z_1 d^2 \rho_2 d z_2
\nonumber \\
 &+&
\sum_{f_1 \ne Q} \int
                 |\Phi^{f_1 \bar f_1}(\rho_1, z_1)|^2
                 |\Phi^{Q \bar Q}(\rho_2, z_2)|^2
         \sigma_{dd}(\rho_1,\rho_2,x_{fQ}) \;
d^2 \rho_1 d z_1 d^2 \rho_2 d z_2 \; ,
\nonumber \\
\label{SM_QQbar}
\end{eqnarray}
where
$\sigma_{dd}$ is the dipole-dipole cross section.

There are two problems associated
with direct use of (\ref{SM_QQbar}).
First of all, it is not completely clear how to generalize
$\sigma_{dd}$ from $\sigma_{dN}$
parametrized in \cite{GW98}.
Secondly, formula (\ref{SM_QQbar}) is correct only at
$W \gg 2 m_Q$. At lower energies one should worry about
proximity of the kinematical threshold.

In a very recent paper \cite{TKM01} a new phenomenological
parametrization for 
$\sigma_{dd}$ has been proposed.
The phenomenological threshold factor in \cite{TKM01}
does not guarantee automatic vanishing of the cross section
exactly below the true kinematical threshold $W = 2 m_a + 2 m_b$.
Therefore, instead of the phenomenological
factor we rather impose an extra kinematical constraint:
$M_{f \bar f} + M_{Q \bar Q} < W$ on the integration
in (\ref{SM_QQbar}).

It is not completely clear how to generalize the energy
dependence of $\sigma_{d N}$ in photon-nucleon scattering
to the energy dependence of $\sigma_{dd}$ in photon-photon
scattering. 
In \cite{szczurek} I have defined the parameter which
controls the SM energy dependence
of $\sigma_{dd}$ in a symmetric way 
with respect to both photons.
In comparison to the prescription in \cite{TKM01}, our prescription leads
to a small reduction of the cross section far from the threshold
\cite{szczurek}.

Up to now we have calculated the contribution when
photons fluctuate into quark-antiquark pairs.
The dipole approach must be supplemented to include the
contribution when either of the photons fluctuates into vector mesons.
If the first photon fluctuates
into the vector mesons, the so-defined single-resolved contribution
to the heavy quark-antiquark production can be calculated
analogously to the photon-nucleon case as
\begin{equation}
\sigma^{SR,1}_{\gamma \gamma \rightarrow Q \bar Q}(W) =
\sum_{V_1} \frac{4 \pi}{f_{V_1}^2} \int
|\Phi_2^{Q \bar Q}(\rho_2,z_2)|^2 \sigma_{V_1 d}(\rho_2,x_1) \;
d^2 \rho_2 dz_2
\; ,
\label{SR_1}
\end{equation}
%
where $\sigma_{V_1 d}$ is vector meson - dipole total cross section.
In the spirit of SM, we parametrize the latter
exactly as for the photon-nucleon case \cite{GW98}
with a simple rescaling of the normalization factor
$\sigma_0^{dV} = 2/3 \cdot \sigma_0^{dN}$.
In the present calculation $\sigma_0^{dN}$ as well as
the other parameters of SM are taken from \cite{GW98}.
Analogously, if the second photon fluctuates into vector mesons
we obtain
\begin{equation}
\sigma^{SR,2}_{\gamma \gamma \rightarrow Q \bar Q}(W) =
\sum_{V_2} \frac{4 \pi}{f_{V_2}^2} \int
|\Phi_1^{Q \bar Q}(\rho_1,z_1)|^2 \sigma_{d V_2}(\rho_1,x_2) \;
d^2 \rho_1 dz_1
\; .
\label{SR_2}
\end{equation}
This clearly doubles the first contribution (\ref{SR_1}) to the total
cross section.

The integrations in (\ref{SR_1}) and (\ref{SR_2}) are not free
of kinematical constraints.
When calculating both single-resolved contributions,
it should be checked additionally if
the heavy quark-antiquark invariant mass $M_{Q \bar Q}$ is
smaller than the total photon-photon energy $W$ (see \cite{szczurek}).

In Fig.2 we show different contributions to the
inclusive $c / \bar c$ (left panel) and $b / \bar b$ (right panel)
production in photon-photon scattering.
The thick solid line represents the sum of all contributions.

Let us start from the discussion of the inclusive charm production.
The experimental data of the L3 collaboration \cite{L3_gg_ccbar}
are shown for comparison. The modifications discussed above
lead to a small damping of the cross section in comparison
to \cite{TKM01}. The corresponding result (long-dashed line)
stays below the recent experimental data of the L3 collaboration
\cite{L3_gg_ccbar}. The hadronic single-resolved contribution
constitutes about 30 - 40 \% of the main SM contribution.
At high energies the cross section for the $2c 2\bar c$ 
component is about 8 \% of that for the single $c \bar c$ pair
component. In the inclusive
cross section its contribution should be doubled because
each of the heavy quarks/antiquarks can be potentially identified
experimentally.

\vspace{-0.1cm}
\begin{figure}
  \begin{center}
\hspace{-1.4cm}
    \includegraphics[width=5.9cm]{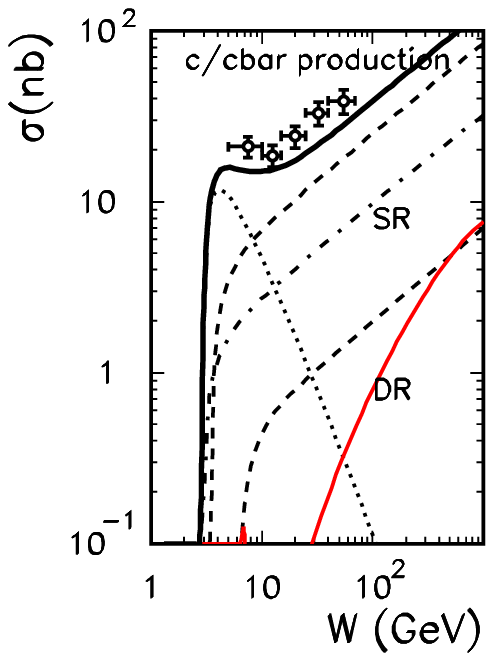}
\hspace{0.0cm}
    \includegraphics[width=5.9cm]{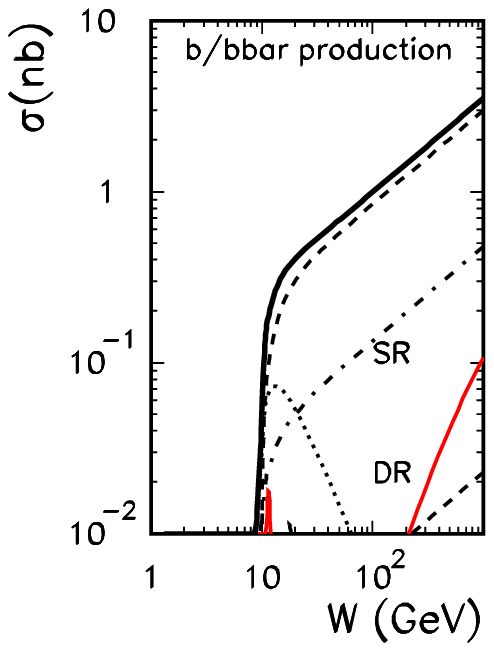}
  \end{center}
\vspace{-1.3cm}
\caption{\small Different contributions to the inclusive charm (left panel)
and bottom (right panel) production.
The long-dashed line:
the dipole-dipole contribution, the dash-dotted line:
the single-resolved contribution,
the lower dashed line: the $2Q  2\bar Q$
contribution, the dotted line: the direct
contribution, the gray solid line: double-resolved contribution.
The experimental data for inclusive $c/\bar c$ production are from
Ref.\cite{L3_gg_ccbar}.
}
\label{figure2}
\end{figure}

At higher energies the direct contribution
is practically negligible.
In contrast, the hadronic double-resolved contribution,
when each of the two
photons fluctuates into a vector meson \cite{szczurek}
is shown by the thin solid line in the figure
becomes important only at very high energies relevant for TESLA.
Here we have consistently taken $g_V(x_V,\mu_F^2) =
g_{\pi}(x_V,\mu_F^2)$.

The situation for bottom production (see right panel) is somewhat
different.
Here the main SM component is dominant. Due to smaller charge
of the bottom quark than that for the charm quark, the direct component
is effectively reduced with respect to the dominant SM component
by the corresponding ratio of quark/antiquark
charges: $(1/9)^2 : (4/9)^2 = 1/16$.
The same is true for the $2 b 2 \bar b$ component.
Here, in addition,
there are threshold effects which play a role up to relatively
high energy.

\section{Conclusions}

There is no common consensus in the literature on detailed
understanding of the dynamics of photon-nucleon and photon-photon
collisions. In this presentation I have limited the discussion
to the production of heavy quarks simultaneously in
photon-nucleon and photon-photon collisions at high energies.
The sizeable mass of charm or bottom quarks sets
a natural low energy limit on a naive application of SM.
Here a careful treatment of the kinematical threshold is required.

We have started the analysis from (real) photon-nucleon scattering,
which is very close to the domain of 
SM as formulated in \cite{GW98}.
If the kinematical threshold corrections are
included, SM gives similar results as the standard
collinear approach for both charm and bottom production.
We have estimated the VDM contribution
to the heavy quark production.

The second part of the present analysis has been devoted to
real photon - real photon collisions.
For the first time in the literature we have estimated
the cross section for the production of $2 c 2 \bar c$ final state.
 We have found that this component
constitutes up to 10-15 \% of the inclusive charm production
at high energies and is negligible for the bottom production.
We have shown how to generalize SM
to the case when one of the photons fluctuates into
light vector mesons. It was found that this component yields
a significant correction of about 30-40 \% for inclusive
charm production and 15-20 \% for bottom production.
We have shown that the double resolved component, when both
photons fluctuate into light vector mesons, is nonnegligible only
at very high energies, both for the charm and bottom production.


\end{document}